\documentclass[12pt]{article}
\usepackage{amssymb,amsmath,bm}

\def\simge{\mathrel{\rlap{\raise 0.511ex \hbox{$>$}}{\lower 0.511ex \hbox{$\sim$}}}}
\def\simle{\mathrel{\rlap{\raise 0.511ex \hbox{$<$}}{\lower 0.511ex \hbox{$\sim$}}}}
\def\slash#1{\setbox0=\hbox{$#1$}\dimen0=\wd0
      \setbox1=\hbox{/} \dimen1=\wd1 \ifdim\dimen0>\dimen1
      \rlap{\hbox to \dimen0{\hfil/\hfil}} #1                        \else
      \rlap{\hbox to \dimen1{\hfil$#1$\hfil}}
      /   \fi}

\begin{document}

\vspace*{-0.5truecm}
\begin{center}
\boldmath

{\Large\textbf{B-Physics: Theoretical Predictions in the LHC Era}}

\unboldmath
\end{center}

\vspace{0.4truecm}

\begin{center}
{\bf Cecilia  Tarantino}
\vspace{0.4truecm}

 {\sl Dipartimento di Fisica, Universit\`a di Roma Tre and INFN Sezione di Roma Tre}

\end{center}
\vspace{0.6cm}
\begin{abstract}
\vspace{0.1cm}
\noindent 
We discuss the status of the theoretical predictions of some interesting b-physics observables that are sensitive to New Physics and can be measured at the LHC.
\end{abstract}\renewcommand{\baselinestretch}{1.2}

\section{Introduction}
The study of flavour physics has allowed to test the Standard Model (SM) with an enormous increase of accuracy, as shown for instance by the few percent uncertainties in the Unitarity Triangle Analysis (UTA).
An even more interesting role of flavour physics that will be extensively explored in the LHC era is the search for New Physics (NP).
The discovery of NP effects in flavour observables, originated by new particles running in loops, in fact, could provide complementary information to collider direct searches. 

Among flavour physics a privileged role is played by b-physics since it presents some advantages in non-perturbative calculations due to the large beauty quark mass ($m_b >> \Lambda_{\rm QCD}$ and therefore $\alpha_s(m_b) \sim 10^{-1}$) and for the wealth of b-observables sensitive to NP.
In particular, the large energy scale $m_b$ implies that soft gluons do not resolve the  spin and flavour of the heavy quark leading to the so-called Heavy Quark Symmetry (HQS), i.e. the invariance of strong interactions under the heavy spin-flavour group $SU(2 N_h)$. The large $m_b$ scale also allows to expand physical quantities in series of $\Lambda_{\rm QCD}/m_b$ through the so-called Heavy Quark Expansion (HQE), where (perturbative) short- and (non-perturbative) long-distance effects are separated.

In what follows we will discuss the status of the theoretical predictions of some of the most interesting b-physics observables that can be accurately measured at the LHC.
These are the $B$-decays that allow to extract the angles of the unitarity triangle (UT), the $B^0_s - \bar B^0_s$ mixing whose rapid oscillations can be measured at the LHC thanks to the large boost, and radiative and rare decays that are highly sensitive to NP.
We will also comment on two recent measurements that represent hot topics as they point towards the presence of NP.
One is the tagged analysis of $B_s \to J_{\Psi} \phi$ by the CDF~\cite{Aaltonen:2007he} and $D \emptyset$~\cite{:2008fj} collaborations that shows a significant deviation of the $B_s$ mixing phase from the SM one~\cite{Bona:2008jn}. The other is the difference between the direct CP-violation in charged $B^{\pm} \to K^{\pm} \pi^0$ decay and the neutral counterpart, which has been measured by the Belle collaboration to be significantly larger than in the SM~\cite{:2008zza}.

\section{Angles of the unitarity triangle}
The UTA has achieved an accuracy of few percent, without revealing any evidence for NP. In particular, a great success of the SM is represented by the good agreement between the measured UT sides and angles.
We note that a precise determination of the CKM parameters is not only an important test of the SM but also a key ingredient to improve the accuracy of the SM predictions in processes sensitive to NP.
In this section we focus on the three angles of the UT, that can all be determined from $B$ decays.

The golden method to determine the angle $\beta$ involves the $B \to J_{\Psi} K_S$ time-dependent CP-asymmetry that, dominated by one tree-level amplitude $b \to c \bar c s$, shows a simple and theoretically clean dependence on $\beta$.
Similar simplifications occur in other $b \to c \bar c s$ channels like $B$-decays to $\psi(2S) K_S$, $\chi_{c1} K_S$, $\eta_c K_S$, $J_{\Psi} K_L$, and $J_{\Psi} K^*$.
There are also determinations of $\beta$ that are sensitive to NP.
They can involve the decays $B \to \eta' K_{S,L}$  and $B \to \phi K_S$, described by the charmless $b \to s$ one-loop (penguin) amplitude.
The ambiguity between the angles $\beta$ and $\pi/2-\beta$, then, has been solved by measuring the positive sign of $\cos (2 \beta)$ from the time-dependent analysis of $B \to J_{\Psi} K^*$ and $B \to D \pi^0$ decays.
On the basis of the measurements mentioned above, the Heavy Flavour Averaging Group (HFAG) quotes the experimental value $\sin (2 \beta)=0.681\pm0.025$~\cite{HFAG}.
From the UTA, excluding the $\beta$ direct measurement, the UTfit collaboration finds $\sin (2 \beta)=0.744\pm0.039$~\cite{UTfit} that is larger than the experimental value, although compatible within the uncertainties.
It has been pointed out that this tension would disappear if instead of the inclusive-exclusive average for $|V_{ub}|$, only the exclusive $|V_{ub}|$ determination was used in the UTA.
The tension, indeed, originates from the difference between $|V_{ub}|_{\rm incl}=(43.1\pm3.9) \cdot 10^{-4}$ and $|V_{ub}|_{\rm excl}=(34.0\pm4.0) \cdot 10^{-4}$, with the inclusive determination affected by a model-dependence in the non-perturbative contribution of the threshold region.  
This explanation is supported by a recent inclusive result $|V_{ub}|_{\rm incl}=(36.9\pm1.3\pm3.1)\cdot10^{-4}$~\cite{Aglietti:2007ik}, lower than the previous inclusive average and in agreement with the exclusive one, obtained with a model based on soft-gluon resummation and an analytic time-like QCD coupling.
More recently, in~\cite{Lunghi:2008aa} the experimental value of $\beta$ has been compared to a $V_{ub}$-independent determination based on the measurements of $|V_{cb}|$, $\Delta M_s/\Delta M_d$ and $\epsilon_K$.
They use for the kaon bag parameter the unquenched lattice result obtained by the RBC collaboration~\cite{Antonio:2007pb} at fixed lattice spacing, $\hat B_K =0.720 \pm 0.013 \pm 0.037$, thus finding $\sin (2 \beta)=0.82 \pm 0.09$, which deviates by around $1.5 \sigma$ from the experimental average.
We believe, however, that before interpreting this deviation as a NP signal, it is important to extrapolate $B_K$ to the continuum limit to further control lattice discretization effects that have been found to be relevant in quenched determinations of $B_K$. 

The angle $\alpha$, related to the phase of $V_{ub}$ ($\alpha \equiv arg[-V_{td} V_{tb}^* / (V_{ud} V_{ub}^*)]$), appears in tree-level $b \to u \bar u d$ transitions which describe charmless B decays like $B \to \pi \pi$, $B \to \rho \rho$, and $B \to \rho \pi$.
These decays receive also a contribution from penguin amplitudes with different CKM factors, that can disentangled by analysing a large set of branching ratios and CP-asymmetries.
The measurements in the $\pi \pi$, $\rho \rho$ and $\rho \pi$ channels have led to the direct determination $\alpha=(91\pm4)^\circ$~\cite{Bona:2005vz} where the main theoretical uncertainty comes from final state interactions and isospin violations in electroweak penguins. 
From the UTA, excluding the $\alpha$ direct measurement, the UTfit collaboration finds $\alpha=(91\pm6)^\circ$~\cite{UTfit}, in perfect agreement with the direct experimental determination.
We note that the region $\alpha \sim 0^\circ$, that would be unphysical as it would imply no CP-violation, is ruled out by simple observations on the hadronic matrix elements~\cite{Bona:2007qta}.
 
The angle $\gamma$, defined by $V_{ub} = |V_{ub}| e^{-i \gamma}$, can be determined from $B \to D K $ decays, where an amplitude interference occurs.
In fact, the decay $B^+ \to D K^+$ can produce both a $D^0$ and a $\bar D^0$, via the tree-level $\bar b \to \bar c u \bar s$ and $\bar b \to \bar u c \bar s$ amplitudes. The $D^0$ and $\bar D^0$ can decay to a common final state and the two corresponding amplitudes interfer with a relative phase $\delta_B \pm \gamma$ (for $B^\pm$), being $\delta_B$ the strong phase.
The irreducible theoretical uncertainty of this procedure comes from the neglection of $D^0-\bar D^0$ mixing effects and is very small, of the order of $0.1$\%.
The methods that have been proposed consider different final states: CP-eigenstates, doubly Cabibbo suppressed $D$ modes or three-body $D$ decaying modes.
The best direct determination of $\gamma$ comes from a combined analysis of many $D$ and $D^*$ modes and yields: $\gamma = (88\pm16)^\circ$~\cite{Bona:2005vz}, compatible with the result $\gamma = (65.1\pm6.5)^\circ$~\cite{UTfit} of the UTA with the exclusion of the direct $\gamma$ measurement.

\boldmath
\section{$B^0_q-\bar B^0_q$ ($q=d,s$) mixing and b-hadron lifetimes} 
\unboldmath
The $B^0_q$-$\bar B^0_q$ system is described by an effective $2 \times 2$ Hamiltonian $\hat H^q = \hat M^q -i \hat \Gamma^q/2$ with the $\hat M^q$ and $\hat \Gamma^q$ matrices describing $B^0_q$-$\bar B^0_q$ oscillations and meson decays, respectively.

The oscillation frequency, given by the meson mass difference $\Delta M_q=2|\hat M_{12}^q|$ between the heavy and light mass eigenstates, requires the calculation of the so-called box-diagrams. The short-distance contribution is well known as the Wilson coefficient has been computed at the NLO in QCD~\cite{Buras:1990fn}.
The operator matrix element involves the decay constant $f_{B_q}$ and the bag parameter $B_{B_q}$, whose uncertainties have been reduced to the ten percent level thanks to recent lattice unquenched calculations (see~\cite{DellaMorte:2007ny} and refs. therein).

The $B_q$ meson lifetimes are calculated, instead, from the diagonal ($\Delta B=0$) elements of the $\hat \Gamma^q$ matrix, while the off-diagonal ($\Delta B=2$) elements provide the width difference $\Delta \Gamma_q$ and the semileptonic CP-asymmetry $A_{SL}^q$.
Both the $\Delta B=0$ and $\Delta B=2$ Wilson coefficients have been computed at the NLO in QCD and at the subleading order in the HQE.
The matrix elements of the leading (in the HQE) operators are known from lattice QCD, while some of the subleading operators are only estimated in the Vacuum Saturation Approximation and represent the main source of theoretical uncertainty.

We now discuss the comparison between theoretical predictions and experimental measurements, paying attention to possible signals of NP.
The $B^0_q-\bar B^0_q$ oscillation frequency has been measured for the first time in 2006 by the CDF collaboration, that found $\Delta M_s = (17.77 \pm 0.10 \pm 0.07)\cdot {\rm ps^{-1}}$~\cite{Abulencia:2006ze} in good agreement with the indirect determination $\Delta M_s = (17.5 \pm 2.1)\cdot {\rm ps^{-1}}$, from the UTA~\cite{UTfit}.
Among b-hadron lifetimes, particular attention goes to the $\Lambda_b$ baryon since, after the new CDF measurement~\cite{Abulencia:2006dr}, the experimental average has increased, leading to the experimental lifetime ratio $\tau(\Lambda_b)/\tau(B_d)=0.904 \pm 0.032$~\cite{HFAG} that is in agreement and slightly larger than the theoretical prediction $\tau(\Lambda_b)/\tau(B_d)=0.88 \pm 0.05$~\cite{Tarantino:2007nf}.
Concerning width differences and semileptonic CP-asymmetries, the theoretical predictions, though including corrections of NLO in QCD and subleading contributions in the HQE, are affected by a theoretical uncertainty of about $30$\% due to strong cancellations between them (see~\cite{Tarantino:2007nf} and refs. therein).
The corresponding experimental measurements still present much larger uncertainties that prevent from significant comparisons with the theoretical predictions. More accurate measurements are certainly looked forward.

The study of $B^0_q$-$\bar B^0_q$ mixing is particularly interesting for its sensitivity to NP. While the $\hat \Gamma^q$ matrix, describing the meson decay to lighter (SM) particles, is expected to be practically insensitive to NP, the $\hat M_{12}^q$ amplitude can receive visible contributions from NP particles running in the box-diagrams.
These effects can be parametrised in terms of a modulus $C_{B_q}$ and a phase $\phi_{B_q}$, respectively equal to one and zero in the SM, by $\hat M_{12}^q=(\hat M_{12}^q)_{\rm SM} C_{B_q} e^{2 i \phi_{B_q}}$. 
We note that Minimal Flavour Violation (MFV) models, which do not introduce new sources of flavour violation in addition to the SM Yukawa couplings, are characterised by $\phi_{B_q}=0$.

For the $B_s$ system a very recent analysis~\cite{Bona:2008jn} finds that $\phi_{B_s}$ deviates by more than $3 \sigma$ from zero, signalling the presence of NP that violates MFV. This analysis combines all the available experimental information on $B_s$  mixing, including the new tagged analysis of $B_s \to J_{\Psi} \phi$ by the CDF~\cite{Bona:2008jn} and D$\emptyset$~\cite{:2008fj} collaborations. While no single measurement has a $3\sigma$ significance yet, all the constraints show a remarkable agreement with the combined result. 
Updated Tevatron analyses and future high-precision measurements at the LHCb will be then of utmost importance. 
We note that, due to the large dominance of the real part of $\Gamma_{12}^s/M_{12}^s$ over the imaginary part, also the semileptonic CP-asymmetry $A_{\rm SL}^s \simeq -Re(\Gamma_{12}^s/M_{12}^s)\,\sin(2\phi_{B_s})/C_{B_s}$ is highly sensitive to $\phi_{B_s}$.
Finally, some examples of NP models where a non-zero value of $\phi_{B_s}$ compatible with the Tevatron measurements is allowed, are the Minimal Supersymmetric Standard Model (MSSM) with a general flavour structure and the Littlest Higgs model with T-parity (LHT) where an enhancement of $A_{\rm SL}^s$ by $10$-$20$ times relative to the SM has been found to be possible~\cite{Blanke:2006sb}.

\section{Radiative and rare decays}
In this section we discuss the radiative and rare $B$ decays $b \to s \gamma$, $B^0_q \to \ell^+ \ell^-$ and $b \to s(d) \ell^+ \ell^-$, whose measurement at the LHC could provide a strong evidence of NP.  

The inclusive decay $B \to X_s \gamma$ is highly sensitive to NP, being a FCNC decay forbidden at tree-level in the SM, and is also theoretically very clean.
The Wilson coefficients have been calculated at the NNLO in QCD and the large logarithms $\log(M_W^2/m_b^2)$ have been resummed using the renormalization group so that the remaining theoretical uncertainty is dominated by unknown $\mathcal{O}(\alpha_s \Lambda_{\rm QCD}/m_b)$ non-perturbative effects in four-fermion operators.
The theoretical prediction reads $Br(B \to X_s \gamma)=(3.15 \pm 0.23) \cdot 10^{-4}$~\cite{Misiak:2006zs} to be compared to the experimental average $Br(B \to X_s \gamma)=(3.55 \pm 0.24) \cdot 10^{-4}$~\cite{HFAG}, with the photon energy cut $E_{\gamma}>1.6$.
The theoretical and experimental results are compatible but a small positive NP effect is still allowed, which significantly constraints NP models.
It represents, for example, the strongest constraint on the one-universal extra-dimension model, yielding the lower bound on the inverse compactification scale $1/R >600\, {\rm GeV}$~\cite{Haisch:2007vb}.
Also in the MSSM the $B \to X_s \gamma$ decay represents an important constraint. Within MFV it points towards a light Higgs mass ($m_h \simeq 110\, {\rm GeV}$)~\cite{Buchmueller:2007zk} compatible with the LEP limit, and beyond MFV it can be translated within the mass insertion approach into a constraint on the non-diagonal entry $(\delta^d_{23})_{LR}$~\cite{Ciuchini:2006dx,Silvestrini:2007yf}.
In LHT it has been found that small ($\sim 4$\%) effects, in the direction of reducing the difference between theoretical and experimental results for $Br(B \to X_s \gamma)$, are possible~\cite{Blanke:2006sb}.   

 The exclusive $b \to s \gamma$ decays are also very sensitive to NP, but theoretically less clean as they require the use of QCD factorization, where the main source of theoretical uncertainty is represented by non-perturbative form factors and light-cone distribution amplitudes.
They are known from lattice QCD~\cite{Becirevic:2006nm} and QCD sum rules on the light cone~\cite{Ball:2006eu}, respectively.
We note that the $B \to K^* \gamma$ decay is theoretically cleaner than $B \to \rho \gamma$ where $\mathcal{O}(\Lambda_{\rm QCD}/m_b)$ corrections turn out to be relevant, and that the t-dependent CP-asymmetries $A_{\rm CP}(B \to K^*(\rho) \gamma)$ are highly sensitive to NP due to their helicity suppression within the SM.
Recently, it has been pointed out that also the $B_s \to \phi \gamma$ channel is very promising~\cite{Muheim:2008vu}.
In the t-dependent $B_s \to \phi \gamma$ decay rate, in fact, the sizable $\Delta \Gamma_s$ width difference allows one to measure the coefficient of the $\sinh (\Delta \Gamma_s /2t)$ term, which is very sensitive to right-handed currents in $B \to V \gamma$ transitions.
 
Among $B_q \to \ell^+ \ell^-$ decays the $\mu^+ \mu^-$ modes are experimentally the most favourable, while the $e^+ e^-$ modes are suppressed by a further factor $m_e^2/m_{\mu}^2$ and the $\tau^+ \tau^-$ have other missing neutrinos from decaying $\tau$'s.
The CDF collaboration has found the experimental upper bounds $Br(B_d\to\mu^+ \mu^-)<1.8\cdot 10^{-8}$ and $Br(B_s\to\mu^+ \mu^-)<5.8\cdot 10^{-8}$ at the $95$\% confidence level~\cite{:2007kv}. 
From the theoretical side these decays are very clean, because leptonic modes are affected by small non-perturbative uncertainties, and highly sensitive to NP, being FCNC processes forbidden at tree-level in the SM and dominated by a penguin operator with Wilson coefficient known at the NLO in QCD.
The SM theoretical predictions read $Br(B_d\to\mu^+ \mu^-)=(1.03\pm0.09)\cdot 10^{-10}$ and $Br(B_s\to\mu^+ \mu^-)=(3.35\pm0.32)\cdot 10^{-9}$~\cite{Blanke:2006ig} and, compared to the present experimental bounds above, show that there is still a lot of room for NP.
A promising way to look for NP in $B_q \to \mu^+ \mu^-$ decays is based on correlations between $Br(B_q\to\mu^+ \mu^-)$ and $\Delta M_q$ and between the $B_s$ and the $B_d$ branching ratios, that in the SM gives $Br(B_s \to \mu^+ \mu^-)/Br(B_d \to \mu^+ \mu^-)=32\pm2$~\cite{Blanke:2006ig}. 
These correlations remain valid in MFV models without new operators in addition to the SM ones.
It will be very interesting to see if the LHC will find some deviations, signalling new operators or new sources of flavour violation w.r.t the SM.
Such deviations could represent a smoking gun for Higgs effects in the  MSSM with large $\tan \beta$, that induce $\tan^6 \beta$ enhanced contributions of new scalar and pseudoscalar operators~\cite{Babu:1999hn,Isidori:2006pk,Huang:1998vb}.

Other FCNC processes that are very sensitive to NP are the rare $b \to s(d) \ell^+ \ell^-$ decays, that in the SM are dominated by the electromagnetic dipole operator $Q_7^{\gamma}$ and the electroweak penguin operators $Q_9$ and $Q_{10}$.
In order to treat the long-distance contribution of $c \bar c$ resonances, two approaches have been used leading to a good control of the hadronic uncertainties of both inclusive and exclusive observables in certain phase-space regions.
The first one consists in introducing appropriate energy cuts, while in the second approach the resonances are modeled and some dispersion relations are used.
Concerning the inclusive $B \to X_s \ell^+ \ell^-$ decays, the Wilson coefficients are known at the NNLO in QCD, and the HQE contributions of order $\Lambda_{\rm QCD}^2/m_c^2$, $\Lambda_{\rm QCD}^2/m_b^2$ and $\Lambda_{\rm QCD}^3/m_b^3$ have been computed as well as QED corrections and bremmstrahlung effects.
The SM theoretical predictions read $Br(B \to X_s \mu^+ \mu^-)=(1.59 \pm 0.11) \cdot 10^{-6}$ and $Br(B \to X_s e^+ e^-)=(1.64 \pm 0.11) \cdot 10^{-6}$~\cite{Huber:2005ig}.
The comparison with future experimental measurements will be of great interest as these branching ratios are sensitive, through the interference between the Wilson coefficients $C_7$ and $C_9$, to NP.
A measurement of the corresponding forward-backward asymmetries would provide complementary information on possible NP effects, being sensitive to the products $C_7 C_{10}$ and $C_9 C_{10}$.
Within the exclusive channels the most promising one is $B \to K^* \ell^+ \ell^-$, and special attention, in the search for NP, is focused on the differential decay widths and the forward-backward and isospin asymmetries.
The SM theoretical predictions for the differential decay widths come from a study in terms of transversity amplitudes~\cite{Lee:2006gs} based on QCD factorization, with the main source uncertainty due to $\mathcal{O}(\Lambda_{\rm QCD}/m_b)$ corrections. NP effects will be visible here if the $C_9$ and $C_{10}$ coefficients turn out to be significantly modified w.r.t. the SM and/or the chirally flipped operators with Wilson coefficients $C'_9$ and $C'_{10}$ are present in the NP model.
A clean way to look for this effects consists in considering the muon to electron ratio where the theoretical uncertainty is enormously reduced up to $0.2$\% and it has been shown in a MFV analysis that NP effects of around $13$\% are possible~\cite{Hiller:2003js}.
In the forward-backward asymmetry and its zero $q_0^2$ the main source of uncertainty is represented by the hadronic inputs that have been obtained from QCD sum rules~\cite{Beneke:2001at} or from the experimental results on $B \to K^* \gamma$~\cite{Ali:2006ew}. A comparison with future measurements will be very useful in constraining NP, in particular the MSSM where significant deviations from the SM are expected~\cite{Ali:1999mm,Feldmann:2002iw}. 
The isospin asymmetry plays a peculiar role as it is very small in the SM (it vanishes in the factorization approximation) and, therefore, could easily reveal a NP contribution.

We conclude this review commenting on a very recent result of the Belle collaboration~\cite{:2008zza} that could point out a signal of NP.
Belle has measured the charged and neutral direct CP-violating decay rate asymmetries, denoted as $A_{K^{\pm}\pi^0}$ and $A_{K^{\pm}\pi^{\mp}}$, respectively. For the charged $B$ meson $A_{K^{\pm}\pi^0}$ is the difference between the number of observed $B^- \to K^- \pi^0$ events versus $B^+ \to K^+ \pi^0$ events, normalized to their sum, and the analogous definition is valid for $A_{K^{\pm}\pi^{\mp}}$ in the neutral case.
The surprising result found by Belle is that the $A_{K^{\pm}\pi^0}$ and $A_{K^{\pm}\pi^{\mp}}$ are measured to be equal to $+7$\% and $-10$\%, respectively, with $A_{K^{\pm}\pi^0}-A_{K^{\pm}\pi^{\mp}}=+0.164\pm0.037$, in contrast with the vanishing result expected in the SM given the current uncertainties.
It will be of great interest to understand the origin of this discrepancy, if this is an evidence of NP or, instead, it reveals that our understanding of strong interaction effects need further clarification.

\vspace*{0.5cm}
I would like to thank the organizers of the {\it V Italian Workshop on the p-p Physics at the LHC} for the invitation to the pleasant workshop held in Perugia.

\end{document}